\documentclass[twocolumn]{aastex62}

\usepackage{cancel}
\usepackage{amsmath}
\usepackage{soul}
\usepackage{longtable}
\bibpunct{(}{)}{;}{a}{}{,}
\usepackage{ae,aecompl}
\usepackage{color,soul}
\usepackage{graphicx}

\usepackage{subfigure}

\usepackage{float}
\usepackage{tikz}
\usetikzlibrary{shapes,arrows}

\usepackage{natbib}

\usepackage{ulem}

\def\acs606{\textit{F606W}}
\def\acs814{\textit{F814W}}

\def\irac36{${\rm 3.6\mu m}$}
\def\irac45{${\rm 4.5\mu m}$}
\def\irac58{${\rm 5.8\mu m}$}
\def\irac80{${\rm 8.0\mu m}$}
\def\mips24{${\rm 24\mu m}$}
\def\pep100{${\rm 100\mu m}$}
\def\pep160{${\rm 160\mu m}$}
\def\her250{${\rm 250\mu m}$}
\def\her350{${\rm 350\mu m}$}
\def\her500{${\rm 500\mu m}$}
\def\irac{{\rm IRAC\/}}
\def\mips{{\rm MIPS\/}}

\def\otelo{\hbox{OTELO}}

\def\acs{\hbox{HST-ACS}}

\def\pep{\hbox{PEP}}

\def\ha{H$\alpha$}
\def\hb{H$\beta$ }

\newcommand{\oii}{[O\,{\sc{ii}}]}
\newcommand{\oiiis}{[O\,{\sc{iii}}]}

\graphicspath{{./}{figures/}}

\received{}
\revised{}
\accepted{}
\submitjournal{ApJL}

\shorttitle{SFR evolution in low-mass galaxies}
\shortauthors{Cedr\'es et al.}

\begin{document}

\title{The OTELO survey: the star formation rate evolution of low-mass galaxies}

\correspondingauthor{Bernab\'e Cedr\'es}
\email{bcedres@iac.es}

\author{Bernab\'e Cedr\'es}
\affil{Instituto de Astrof\'isica de Canarias (IAC), E-38200 La Laguna, Tenerife, Spain}
\affil{Departamento de Astrof{\'i}sica, Universidad de La Laguna (ULL), 38205 La Laguna, Tenerife, Spain}

\author{Ana Mar{\'i}a P\'erez-Garc{\'i}a}
\affiliation{ Centro de Astrobiolog{\'i}a (CSIC/INTA), 28692 ESAC Campus, Villanueva de la Cañada, Madrid, Spain}
\affiliation{Asociaci\'on Astrof{\'i}sica para la Promoci\'on de la Investigaci\'on, Instrumentaci\'on y su Desarrollo, ASPID, 38205 La Laguna, Tenerife, Spain}

\author{Ricardo P\'erez-Mart\'inez}
\affiliation{ISDEFE for European Space Astronomy Centre (ESAC)/ESA, P.O. Box 78, E-28690 Villanueva de la Ca\~nada, Madrid, Spain}
\affiliation{Asociaci\'on Astrof{\'i}sica para la Promoci\'on de la Investigaci\'on, Instrumentaci\'on y su Desarrollo, ASPID, 38205 La Laguna, Tenerife, Spain}

\author{Miguel Cervi\~no}
\affiliation{ Centro de Astrobiolog{\'i}a (CSIC/INTA), 28692 ESAC Campus, Villanueva de la Cañada, Madrid, Spain}

\author{Jes\'us Gallego}
\affiliation{Departamento de F\'isica de la Tierra y Astrof\'isica, Instituto de F\'isica de Part\'iculas y del Cosmos, IPARCOS. Universidad Complutense de Madrid, E-28040, Madrid, Spain}

\author{\'Angel Bongiovanni}
\affiliation{Institut de Radioastronomie Millim\'etrique (IRAM), Av. Divina Pastora 7, N\'ucleo Central 18012, Granada, Spain}
\affiliation{Asociaci\'on Astrof{\'i}sica para la Promoci\'on de la Investigaci\'on, Instrumentaci\'on y su Desarrollo, ASPID, 38205 La Laguna, Tenerife, Spain}

\author{Jordi Cepa}
\affil{Instituto de Astrof\'isica de Canarias (IAC), E-38200 La Laguna, Tenerife, Spain}
\affil{Departamento de Astrof{\'i}sica, Universidad de La Laguna (ULL), 38205 La Laguna, Tenerife, Spain}
\affiliation{Asociaci\'on Astrof{\'i}sica para la Promoci\'on de la Investigaci\'on, Instrumentaci\'on y su Desarrollo, ASPID, 38205 La Laguna, Tenerife, Spain}

\author{Roc\'io Navarro Mart\'inez}
\affiliation{Asociaci\'on Astrof{\'i}sica para la Promoci\'on de la Investigaci\'on, Instrumentaci\'on y su Desarrollo, ASPID, 38205 La Laguna, Tenerife, Spain}

\author{Jakub Nadolny}
\affil{Instituto de Astrof\'isica de Canarias (IAC), E-38200 La Laguna, Tenerife, Spain}
\affil{Departamento de Astrof{\'i}sica, Universidad de La Laguna (ULL), 38205 La Laguna, Tenerife, Spain}

\author{Maritza A. Lara-L\'opez}
\affil{Armagh Observatory and Planetarium, College Hill, Armagh BT61 9DG, Northern Ireland, UK}

\author{Miguel S\'anchez-Portal}
\affiliation{Institut de Radioastronomie Millim\'etrique (IRAM), Av. Divina Pastora 7, N\'ucleo Central 18012, Granada, Spain}
\affiliation{Asociaci\'on Astrof{\'i}sica para la Promoci\'on de la Investigaci\'on, Instrumentaci\'on y su Desarrollo, ASPID, 38205 La Laguna, Tenerife, Spain}

\author{Emilio J. Alfaro}
\affil{Instituto de Astrof\'isica de Andaluc\'ia, CSIC, E-18080, Granada, Spain}

\author{Jos\'e A. de Diego}
\affil{Instituto de Astronom\'ia, Universidad Nacional Aut\'onoma de M\'exico, Apdo. Postal 70-264, 04510 Ciudad de M\'exico, Mexico}

\author{Mauro Gonz\'alez-Otero}
\affil{Instituto de Astrof\'isica de Canarias (IAC), E-38200 La Laguna, Tenerife, Spain}
\affil{Departamento de Astrof{\'i}sica, Universidad de La Laguna (ULL), 38205 La Laguna, Tenerife, Spain}
\affiliation{Asociaci\'on Astrof{\'i}sica para la Promoci\'on de la Investigaci\'on, Instrumentaci\'on y su Desarrollo, ASPID,\\ 38205 La Laguna, Tenerife, Spain}

\author{J. Jes\'us Gonz\'alez}
\affil{Instituto de Astronom\'ia, Universidad Nacional Aut\'onoma de M\'exico, Apdo. Postal 70-264, 04510 Ciudad de M\'exico, Mexico}

\author{J. Ignacio Gonz\'alez-Serrano}
\affil{Instituto de F\'isica de Cantabria (CSIC-Universidad de Cantabria), E-39005 Santander, Spain}

\author{Carmen P. Padilla Torres}
\affil{INAF, Telescopio Nazionale Galileo, Apartado de Correos 565, E-38700 Santa Cruz de la Palma, Spain}
\affil{Fundaci\'on Galileo Galilei - INAF Rambla Jos\'e Ana Fern\'andez P\'erez, 7, 38712 Bre\~na Baja,
Tenerife, Spain}
\affiliation{Asociaci\'on Astrof{\'i}sica para la Promoci\'on de la Investigaci\'on, Instrumentaci\'on y su Desarrollo, ASPID, 38205 La Laguna, Tenerife, Spain}




\begin{abstract}
We present the analysis of a sample of \ha\,, \hb\ and \oii\ emission line galaxies from the \otelo\ survey, with masses typically below $log(M_*/M_\sun) \sim 9.4$ and redshifts between $z \sim 0.4$ and 1.43.
We study the star formation rate, star formation rate density, and number density and their evolution with redshift.
We obtain a robust estimate of the specific star formation rate -- stellar mass relation based on the lowest mass sample published so far. 
We also determine a flat trend of the star formation rate density and number density with redshift.
Our results suggest a scenario of no evolution of the number density of galaxies, regardless of their masses, up to redshift $z\sim1.4$. This implies a gradual change of the relative importance of the star forming processes, from high-mass galaxies to low-mass galaxies, with decreasing redshift. We also find little or no variation of the star formation rate density in the redshift range of $0.4<z<1.43$.
\end{abstract}



\section{Introduction} \label{sec:intro}

The study of star formation along the cosmic times provides an outstanding insight into the main physical processes driving the evolution of galaxies at cosmological scales. 
There is 
a tight relationship between SFR and stellar mass for Star Forming Galaxies --the Main Sequence (MS)-- in both local and high redshift universe \citep[][and many others]{Brichmann04}.
Focusing on the low mass range,  the study of the Star Formation Rate (SFR) of dwarf galaxies allows establishing a direct connection to the early epochs of galaxy formation.
\citet{Whitaker2014} found that the evolution of low-mass galaxies is steeper than in massive galaxies for galaxies with masses larger than $10^9\, \mathrm{M}_\sun$, using SFR obtained from UV and FIR continuum. In the same way, \citet{Gonzalez14}  have shown that for sources with $M_* \sim 5 \times 10^9\, \mathrm{M}_\sun$ the specific Star Formation Rate (sSFR) shows no evidence of significant evolution from $z\sim 2$ to $z \sim 7$ ($\mathrm{sSFR} \sim  2\, \mathrm{Gyr}^{-1}$). On the other hand, results of the EAGLE and Illustris  simulations suggest that low mass galaxies behave differently than more massive ones, agreeing with observations for masses larger than  $10^9\, \mathrm{M}_\sun$ \citep{Schaye15,Vogelsberger14}.     \\

Previous works have shown that the SFR density (SFRD) declines along cosmic time \citep[see, for example][]{harish2020,sobral2013}, although the samples used are composed by high mass galaxies. Some studies, as \citet{Mobasher09}, with a sample of galaxies with masses larger than $10^{9.5}\, \mathrm{M}_\sun$, concludes that the massive systems have had their major star-formation activity at earlier epochs ($z>2$) than the lower-mass galaxies.  \citet{davies2009}, analyzing a sample of dwarf galaxies at $z\sim$1, suggests that for low-mass galaxies, the SFRD is roughly constant from $z=1$ to now. However, there are no conclusive studies for low mass galaxies at  non-local redshifts, mainly because of the difficulty of detecting such faint objects.

In this letter, we present the results of the analysis of the star formation activity of a sample of galaxies with masses as low as $\log(M_*/M_\sun) \sim 7.12$, with 80\% of them below $\log(M_*/M_\sun) \sim 9.4$ at redshifts $\sim$ 0.4 to 1.4 from the OTELO survey, with the purpose to establish the evolution of the SFR and SFRD for these low mass objects.

Through this paper we assume a standard $\Lambda$-cold dark matter cosmology with $\Omega_\Lambda = 0.7$, $\Omega_m = 0.3$, and $\mathrm{H}_0 = 70 \,\mathrm{km}\,  \mathrm{s}^{-1}\, \mathrm{Mpc}^{-1}$.

\section{The OTELO samples}\label{sec:otelo}

This work takes advantage of the \otelo\ survey and its data-products \citep{bongiovanni2019}. The \otelo\ survey is a narrow-band scan ultra deep pencil-beam survey carried out with tunable filters of the OSIRIS instrument on the Gran Telescopio Canarias. It covers a region of 56 arcmin$^2$ of the Extended Groth Field. Observations through 36 tunable filter images in the $8950 - 9300$\,\AA ~region provides a  pseudospectra ($\mathrm{R}\sim 700$) from which emission lines can be identified and measured. The limiting flux detected in the OTELO survey is $\sim$1. 10$^{-19}$ erg/cm$^2$/s/\AA\ which constitutes the deepest narrow-band survey up to date. These data are complemented with $u$, $g$, $r$, $i$, $z$, $J$, $H$ and $K_s$ photometry.
Redshift and extinction estimates has been obtained from \textit{Le Phare} code (\citealt{lephare2006}) using representative templates of Hubble types from \cite{coleman1980}, and starburst galaxies from \cite{kinney1996} and different $E(B-V)$ assuming \cite{calzetti2000} law. The estimations of $E(B-V)$ values were sampled by the code in intervals of 0.05\,mag. We obtained median extinctions $E(B-V)\simeq0.1$, $E(B-V)\simeq$0.1 and $E(B-V)\simeq$0.15 for \ha, \hb\ and \oii\ respectively (see \citealt{RamonPerez2019b}, \citealt{rocio2021} and \citealt{beli2021}).

A direct consequence of the design of the \otelo\ survey is the favoring of detection of low-mass galaxies ($M_*<10^{10}M_{\sun}$), which compose more than the 87\% of the total detected emitters (see, for example, \citealt{beli2021} and \citealt{bongio2020}).

The Luminosity Functions (LFs) covering the main emission lines for the \otelo\ 
emitters has already been calculated in 
\cite{RamonPerez2019b}, \cite{rocio2021} and \cite{beli2021} for redshifts $z\sim0.4$ (\ha), $z\sim0.9$ (\hb) and $z\sim1.43$ (\oii{}), respectively. An in-depth description of the methods and corrections applied (including the corrections from AGNs presence, completeness and cosmic variance) can be found in those works.

In Table \ref{param}, a summary of the data employed in this study is presented. This table includes the number of emitters, the volume sampled at each band observed, and the \cite{Schechter76} parameters for the LF fit.

\begin{deluxetable*}{lccccccc}

    \tablecaption{Data sample description and Luminosity Functions\label{param}}
    \tablewidth{2pt}
    \tablehead{
        \colhead{Employed Emission line} & \colhead{Number of} & \colhead{$\langle z \rangle$} & \colhead{$\langle$V$_c$ $\rangle$
} & \colhead{$\log \phi^{*}$} & \colhead{$\log L^*$} & \colhead{$\alpha$}  \\
        \colhead{} & \colhead{emitters} & \colhead{} & $10^3\,\mathrm{Mpc}^3$ & \colhead{[${\rm Mpc}^{-3}{\rm dex}^{-1}$]} & \colhead{[$\mathrm{erg}\,\mathrm{s}^{-1}$]} &\colhead{}  }
    \startdata
        \ha\, (\citealt{RamonPerez2019b})  & 46 & 0.40 & 1.40 &  -2.75$\pm$0.19 & 41.85$\pm$0.22 & -1.21$\pm$0.07 \\
        \hb\, (\citealt{rocio2021})  & 40 & 0.90 & 5.19 & -2.77$\pm$0.12 & 41.34$\pm$0.27 & -1.43$\pm$0.10  \\
        \oii\, (\citealt{beli2021})  & 60 & 1.43 & 10.21 & -3.23$\pm$0.11 & 42.44$\pm$0.11 & -1.41$\pm$0.09  \\
        \tableline
   \enddata
\end{deluxetable*}

\section{SFR and sSFR}\label{sec:sfr}
The SFR was calculated following \cite{Kennicutt1998}, employing a \cite{kroupa2001} initial mass function (IMF).
\begin{equation}
    \mathrm{SFR}(\mathrm{M_\odot\,yr^{-1})}=5.29\times10^{-42}L(\mathrm{H}\alpha)\mathrm{(erg\,s^{-1})},
    \label{eq:sfrha}
\end{equation}
\begin{equation}
    \mathrm{SFR}(\mathrm{M_\odot\,yr^{-1})}=1.507\times10^{-41}L(\mathrm{H}\beta)\mathrm{(erg\,s^{-1})},
    \label{eq:sfrhb}
\end{equation}
for \ha\ and \hb respectively. For \oii, due to its dependence on metallicity, we choose to use the improved \cite{Kennicutt1998} calibration proposed by \cite{kewley2002}, 
\begin{equation}
    \mathrm{SFR}(\mathrm{M_\odot\,yr^{-1})}=\frac{5.29\times10^{-42}L([\mathrm{O}\,
    \textsc{ii}])(\mathrm{erg\,s^{-1})}}{-1.75\times[\log(\mathrm{O/H})+12]+16.73},
    \label{eq:sfroii}
\end{equation}
where $\log(\mathrm{O/H})+12$ is the oxygen abundance for each of the \oii\ emitters. Based on \cite{henry2013}, we assumed a mean value for the metallicity for the gas phase of dwarf galaxies at our redshift of $\sim8.5$. To derive the uncertainties due to the use of this mean value of the metallicity instead of a properly derived quantity for each emitter, we assumed that the real abundance could vary from a minimum of $\sim8.2$ to a maximum of $\sim8.8$ (values given in \citealt{henry2013}). Then, we calculated the values for the SFR for the minimum and maximum metallicity and used them as extreme values for the SFR. The differences between the extreme SFRs and the SFR from the mean oxygen abundance gives us the uncertainty range in the SFR due to the metallicity dependence.

 The \textit{Le Phare} code (see section \ref{sec:otelo}) does not provide an evaluation of the errors in $E(B-V)$, therefore the uncertainty in this term was no taken into account when determining the uncertainty in the SFR. Nevertheless, the derived values of $E(B-V)$ are small so the effect of dust extinction on the derived SFRs should be small, although it would introduce some extra scatter in our SFR estimates. It should also be noted that taking into account that all the lines employed in this study sample a similar region of the ionizing continuum, and hence the same time scale of the star formation activity (see for example \citealt{miguel2016} and references therein), we can assume that all SFRs derived can be compared with each other.

The stellar mass for each emitter was derived following \cite{ls2019} 
for quiescent and SFGs, employing rest-frame g- and i-band magnitudes, together with absolute i-band magnitude
(see \citealt{jakub2020} for details).
The range of the masses derived were  $10^{7.1}<{\rm M_{*}/M_{\sun}}<10^{9.6}$ for \ha\, $10^{7.6}<{\rm M_{*}/M_{\sun}}<10^{10.7}$ for \hb\ and $10^{7.89}<{\rm M_{*}/M_{\sun}}<10^{10.93}$ for \oii.

\begin{figure*}[ht]
 \gridline{\fig{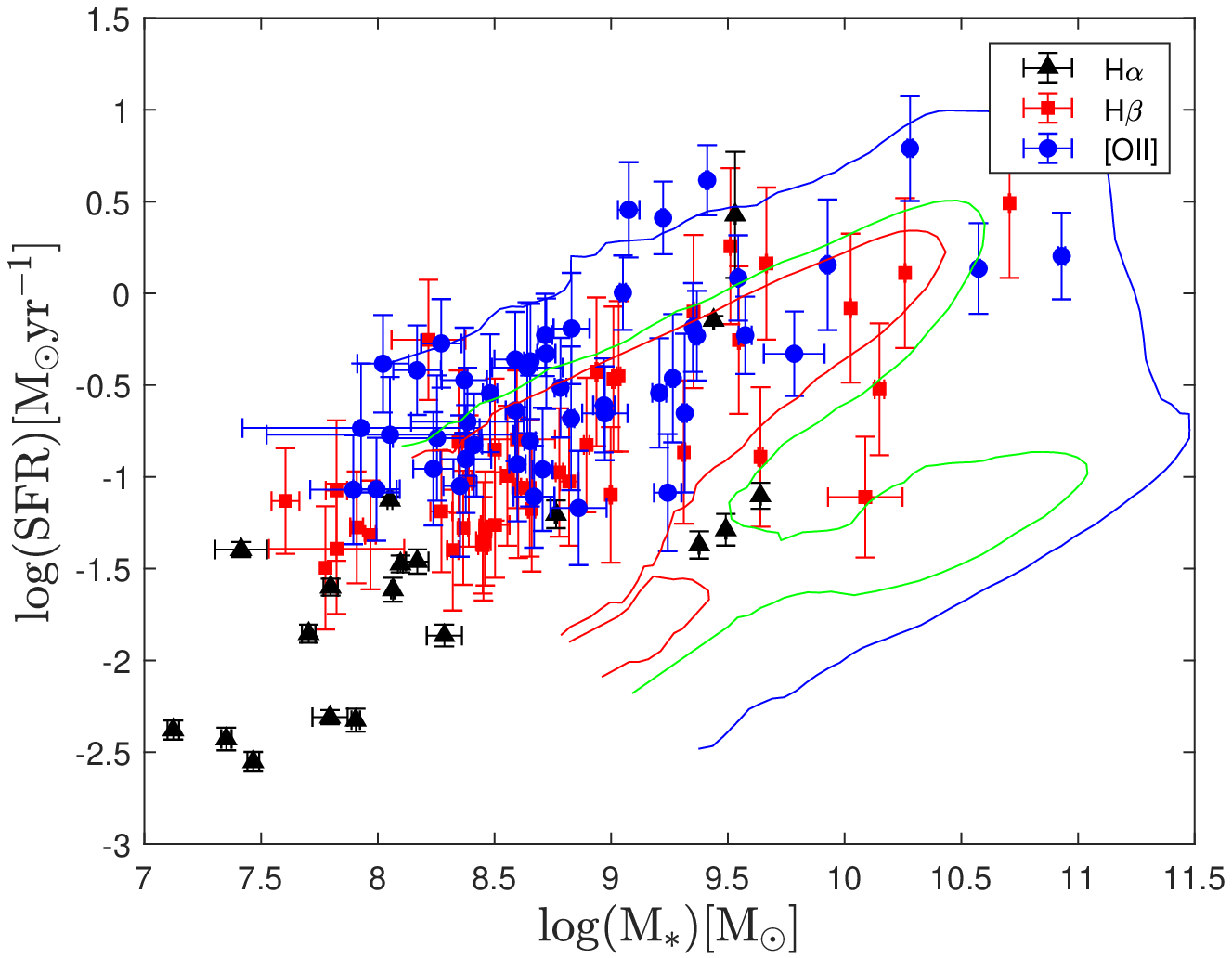}{0.5\textwidth}{(a)}
 \fig{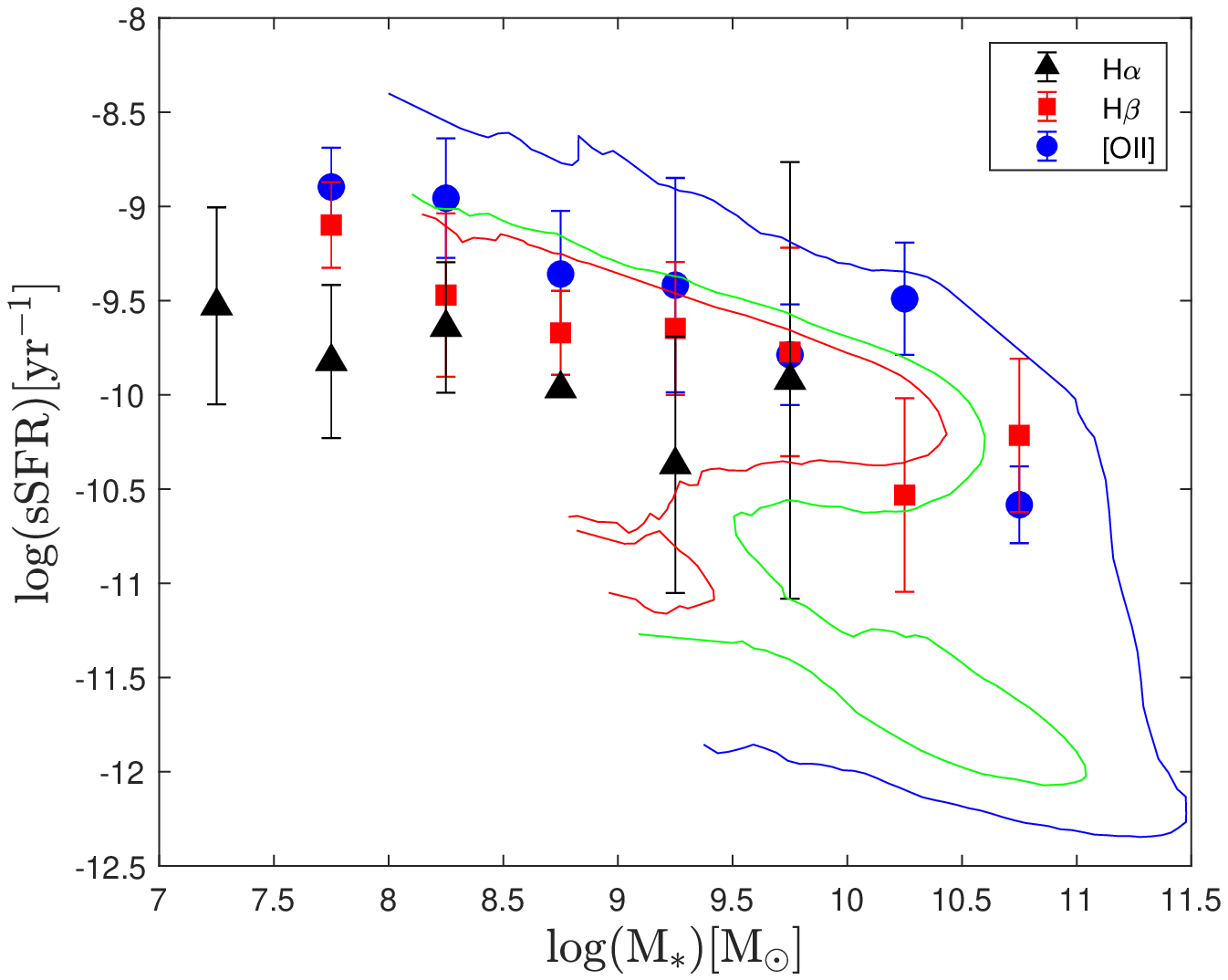}{0.5\textwidth}{(b)}}
 \caption{Logarithm of the SFR  (panel a) and the stacked sSFR (panel b) as a function of the logarithm of the stellar mass. Black triangles are the data from \ha\ (\citealt{RamonPerez2019b}), red squares are the data from \hb\ (\citealt{rocio2021}) and blue circles are the data from \oii\ emitters (\citealt{beli2021}). Contours corresponds to the number of galaxies from the SDSS database obtained by \cite{RP2015} at values of  $2.0 \times 10^4$\,(blue), $7.0\times 10^4$\, (green) and $1.2\times 10^5$\,(red),
 being this last one the definition of the local SF-Main Sequence .}
 \label{sfrpanel}
\end{figure*}

In Fig. \ref{sfrpanel} panel (a) we have represented the SFR as a function of the stellar mass of the galaxies. 
A small difference in the SFR with redshift may exist, with the \ha\ detected galaxies having lower values of the SFR when compared with the galaxies detected in \hb\ and \oii. This difference may be attributed to a selection effect of the sources. Indeed, the mean value of the SFR in \ha\ is $0.125\, \mathrm{M}_{\odot}\,\mathrm{yr}^{-1}$, so if we want to observe such SFR in the \oii\ line, the value of the  luminosity has to be $2.36\times10^{40}\,\mathrm{erg/s}$. However, the minimum value in luminosity detected for the \oii\ galaxy sample is $2.37 \times10^{40}\,\mathrm{erg/s}$ (see \citealt{beli2021}). 
Nevertheless, taking into account the uncertainties, the results for \ha\ are compatible with the rest of the lines (\hb\ and \oii) in the mass range in which they coincide.

In Fig. \ref{sfrpanel}, panel (b) we have represented the logarithm of the specific star formation rate (sSFR) as a function of the logarithm of the stellar mass, stacked in bins of 0.5 dex in $\log (M_*)[\mathrm{M}_\odot]$.
Although it seems that the results for the \oii\ emitters present a larger value for the sSFR when compared with the \ha\ and \hb\ emitters, if we take into account the uncertainties, this figure suggests that the bulk of all the emitters, regardless of the observed line, are located inside the area of the MS defined by the number of galaxies from the SDSS database (\citealt{RP2015}).

Following \cite{gilbank2011}, we may assume a power-law relationship between the sSFR and the stellar mass, such as $\mathrm{sSFR}\propto M_{*}^{\beta}$. Our data sample have masses down to $\log(M_*/M_{\odot})\sim7.12$ for \ha\,, and median values of $\log(M_*/M_{\odot})=8.06$, $\log(M_*/M_{\odot})=8.62$ and $\log(M_*/M_{\odot})=8.72$,  with 80\% of galaxies below the masses $\log(M_*/M_{\odot})=9.4$, $\log(M_*/M_{\odot})=9.5$ and $\log(M_*/M_{\odot})=9.3$ for \ha\ , \hb\ and \oii\ respectively. We obtain a least-squared fit to these stacked data with $\beta_{\rm H\alpha}=-0.1\pm0.2$, $\beta_{\rm H\beta}=-0.5\pm0.2$ and $\beta_{\rm [OII]}=-0.6\pm0.3$.

This is consistent with previous works performed with higher mass ranges. \cite{ramraj2017} obtained the sSFR for the \ha\ line at $z\sim1$ in the mass range $8.5<\log(M_*/M_{\odot})<9.5$, and a value of $\beta=-0.47\pm0.04$, and also from the data obtained by \cite{gilbank2010}, a value of $\beta=-0.08\pm0.01$ for $z\sim0.1$ in that mass range.

The depth of the data used is crucial to determine the slope of the sSFR-stellar mass relation, as also noted in \citet{ramraj2017}. The \otelo\ survey is able to sample the lowest mass ranges published to date for the objects we study in this letter, allowing us to provide the most robust estimation of the steepness of this relation. 
Our results reinforce the idea of flatter slopes for $\beta$ at intermediate redshifts than the previous ones presented in literature (see for example \citealt{sobral2011}, where they obtained $\beta=-1.0\pm0.07$ at $z\sim1$). 

\section{SFR density and number density evolution}
\label{sec:sfrd_dens}

\subsection{SFR density}
\label{subsec:sfrd}
As explained in section \ref{sec:otelo}, the LF has been obtained for each band employed in this study.
 This LF may be integrated to obtain the total luminosity density. According with \cite{Schechter76}, the integral is equal to:
\begin{equation}
    \mathcal{L}=\int_{0}^{\infty}\phi(L)L\,{\rm d}L=\phi^{\star}L^{\star}\,\Gamma(\alpha+2),
    \label{eq:lumfull}
\end{equation}
\noindent where $\Gamma$ is the gamma function.

This integrated luminosity density can be converted into star formation rate density (SFRD) by simply substituting the normal luminosity $L$ by $\mathcal{L}$ in equations \ref{eq:sfrha}, \ref{eq:sfrhb} and \ref{eq:sfroii} for \ha, \hb\ and \oii\ respectively. 

\begin{figure*}[ht]
 \gridline{\fig{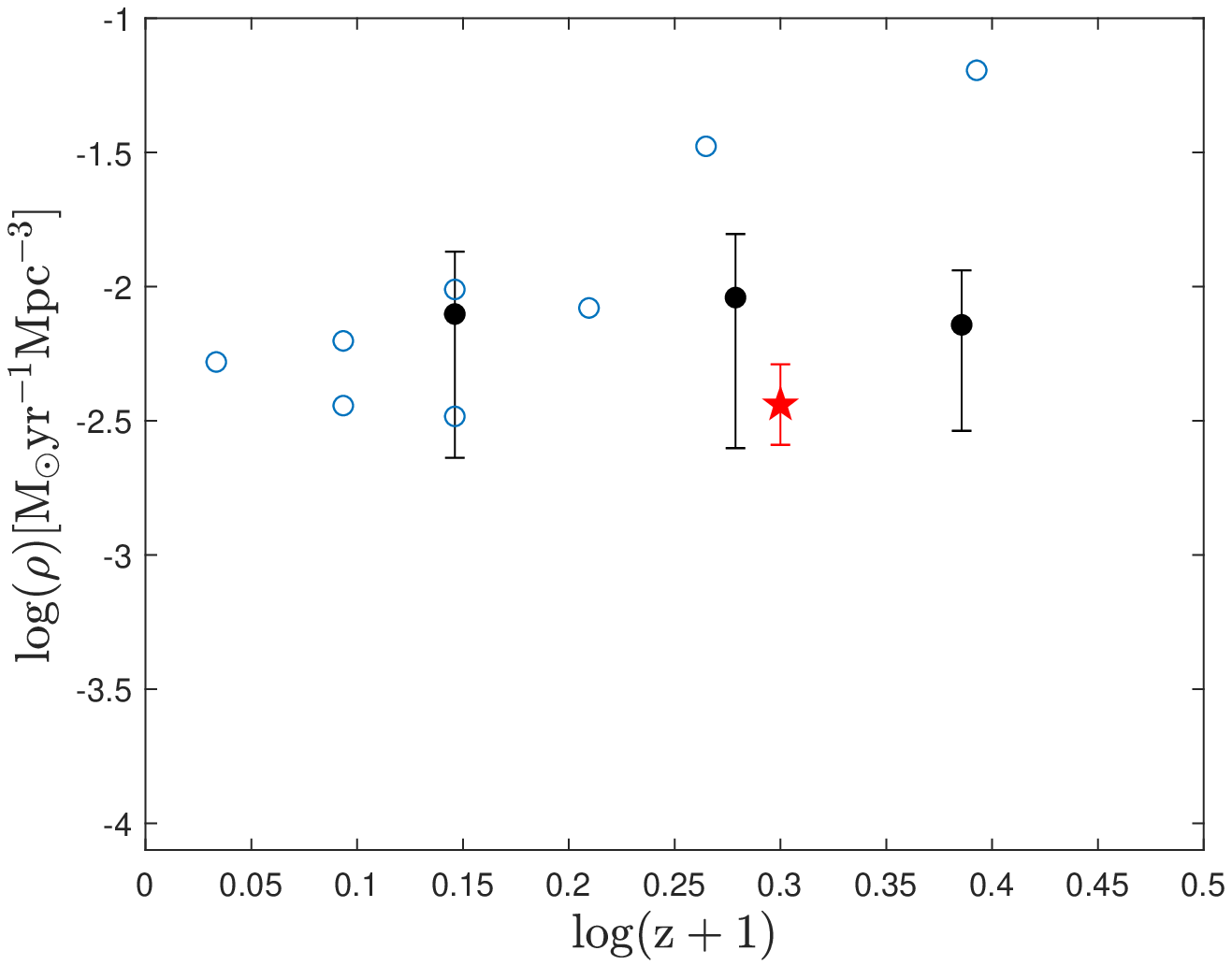}{0.5\textwidth}{(a)}
 \fig{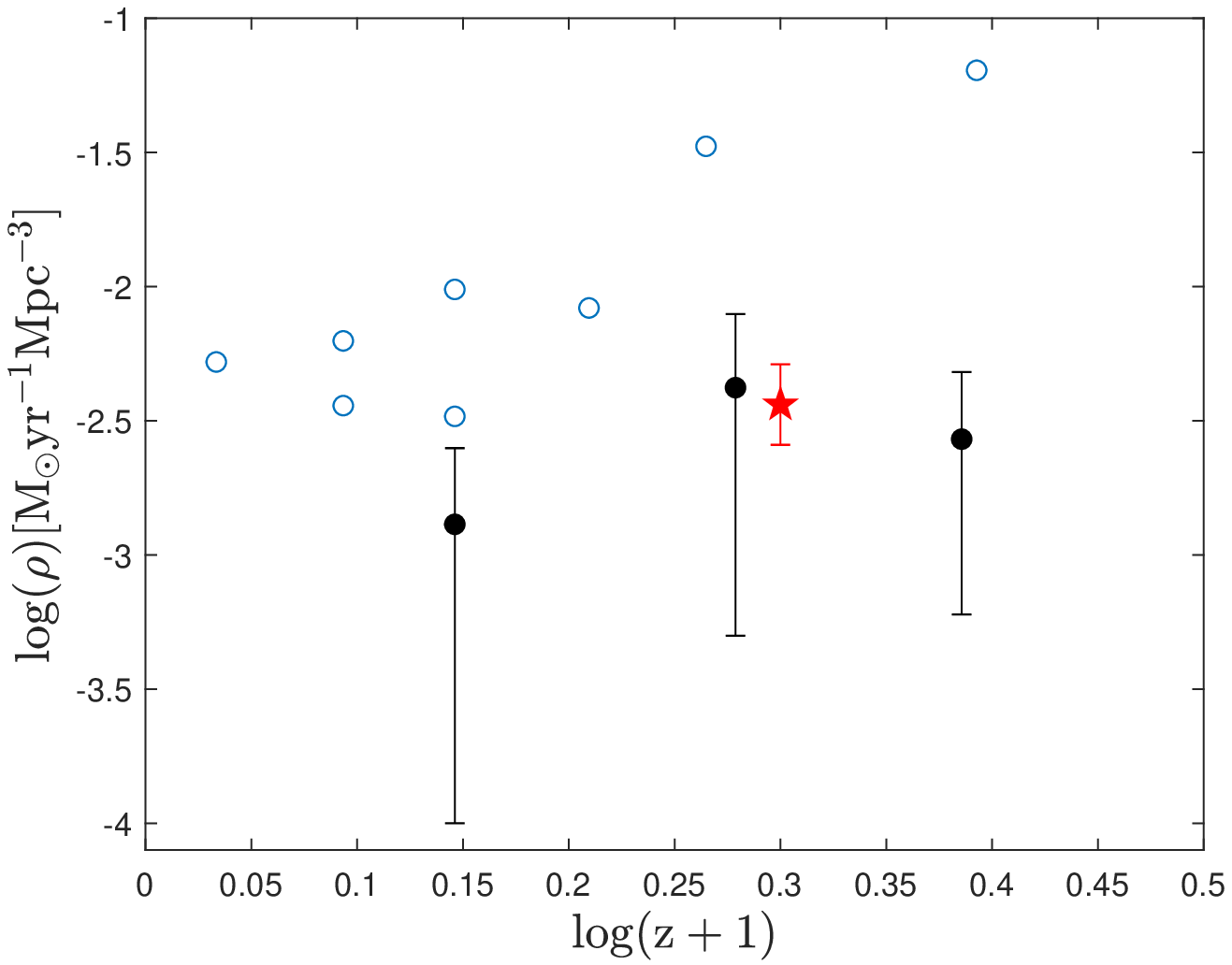}{0.5\textwidth}{(b)}}
 \caption{Star formation density as a function of $\log(z+1)$. Our results are the black dots. Blue circles are the data from \cite{ly2007},  \cite{morioka2008}, \cite{harish2020} and \cite{sobral2013}, the red star is the result from only low mass galaxies extracted from \cite{davies2009}. The SFRD for all the data is calculated employing \cite{kroupa2001} IMF. In panel (a) the integration is done up to infinity in mass, in panel (b) is done up to $M_{*}<10^{9}\, {\rm M_{\odot}}$.}
 \label{sfhpanel}
\end{figure*}

In Fig.\ref{sfhpanel} panel (a) we have represented the SFRD as a function of the redshift. Our data are the black dots (\ha, \hb and \oii), the open cyan circles are data from \cite{ly2007}, \cite{morioka2008}, \cite{harish2020} and \cite{sobral2013}, all derived from the integration of their proposed LFs. The red star is the data for dwarf galaxies only from \cite{davies2009}. It is clear from the \otelo\ data that there is little to no evolution of the SFRD for low-mass galaxies. Moreover, \otelo\ results seem to be closer to the one obtained by \cite{davies2009}. In \cite{beli2021} is discussed that part of the low value for the SFRD in \oii\ may be explained by the large uncertainty presented in $\phi^*$ due to the cosmic variance (CV). The same may happen with the other lines. Nevertheless, this effect may include uncertainty on $\phi^*$ of about 100\%, which has been taken into account when deriving the errorbars in the SFRD. Moreover, the uncertainty due to the CV will have a random effect in the final value of the SFRD, but in our case, all the lines present the same behavior, which makes it compatible with a no evolution scenario.

On the other hand, there is an increase in the SFRD with redshift for samples containing low and high-mass galaxies. For low redshift, there is an agreement between \otelo\ and the data from literature (\citealt{ly2007} for emitters at $z\sim0.07$, \citealt{morioka2008} for emitters at $z\sim0.24$ or \citealt{sobral2013} for emitters at $z\sim0.4$). This confirms the suggestion of \cite{davies2009}, where it is pointed out that the bulk of the SFRD shifts from high-mass to low-mass galaxies with decreasing redshift. Meanwhile, the SFRD in low-mass galaxies has remained fairly constant with redshift, and it is a clear indication of the \textit{downsizing} first suggested in \cite{cowie1996} and detected in \cite{weisz2011} or in \cite{rodriguez-munoz2015}, where it is hinted that at least 90\% of the stellar mass for low-mass galaxies is preferably formed at low redshift ($z\lesssim0.15$).

In Fig. \ref{sfhpanel} panel (b) we have represented the SFRD for \otelo\ survey integrated up to $M_{*}=10^{9}\, {\rm M_{\odot}}$ (black dots), to include only low-mass galaxies. This was done in order to better compare our results with that from \cite{davies2009} (represented as a red star). Indeed, the main body of our data comes from low-mass galaxies, but the integration of the LF from equation \ref{eq:lumfull} is up to infinity, which includes some contribution by high-mass galaxies that may bias the value of the derived SFRD.

We can obtain the contribution to the luminosity density of only the low-mass galaxies analytically by doing the following:
\begin{equation}
\begin{split}
    \mathcal{L}_{\mathrm{low-mass}}=\mathcal{L}_{\mathrm{total}}-\mathcal{L}_{\mathrm{high-mass}}=\\
    =\phi^*L^*\left[\Gamma(\alpha+2)-\Gamma(\alpha+2,L/L^*)\right] ,
\end{split}
\label{eq:lumlowmass}
\end{equation}
where $\Gamma(\alpha+2,L/L^*)$ is the incomplete gamma function, and $L$ is the maximum value for the luminosity to integrate the luminosity density, corresponding to the luminosity of the maximum value of the stellar mass, in this case $M_{*}=10^{9}\, {\rm M_{\odot}}$. To estimate the limiting value of $L$, we assumed a linear relationship between the continuum of the emitters at the observed bands and the stellar mass, and obtained a limiting $L_{cont}$ for the emitters with $M_{*} \leq 10^{9}\, {\rm M_{\odot}}$.
We also assumed that the equivalent width of the lines is approximately constant for low masses up to $M_{*}=10^{9}\, {\rm M_{\odot}}$. We know that this is a crude approximation, but the assumption of a constant equivalent width is roughly valid for our sample, as it is shown in Fig. 9 of \cite{beli2021} for the case of \oii\ emitters.
Employing this, we were able to determine, in a somewhat crude way, a limiting value for the luminosity of the line $L$ from equation \ref{eq:lumlowmass}. The obtained  limits were $\log(L[\mathrm{erg\, s^{-1}}])=40.26$, 40.71 and 41.27  for \ha, \hb\ and \oii\ respectively. With those values, the resulting $\mathcal{L}_{\mathrm{low-mass}}$ from equation \ref{eq:lumlowmass} is then converted in SFRD following equations \ref{eq:sfrha}, \ref{eq:sfrhb} and \ref{eq:sfroii}.

From Fig. \ref{sfhpanel} panel (b) it is now clearer the agreement of our values of the SFRD with the results from \cite{davies2009}. However, it seems that the \ha\ emitters may present a lower value of the SFRD when compared with the other emitters, even if the general result is well inside the derived uncertainties. This can be explained by the very low sampled volume for the \ha\ emitters (see table \ref{param}), which is 5 and 10 times smaller in this line than in \hb\ and \oii\ respectively. This implies a larger uncertainty due to the cosmic variance (\citealt{stroe2015}) when compared with \hb\ and \oii. Even so, it seems that the non-evolutionary scenario is preserved.  

\subsection{Number density}

\begin{figure}[ht]
   \centering
   \includegraphics[width=\hsize]{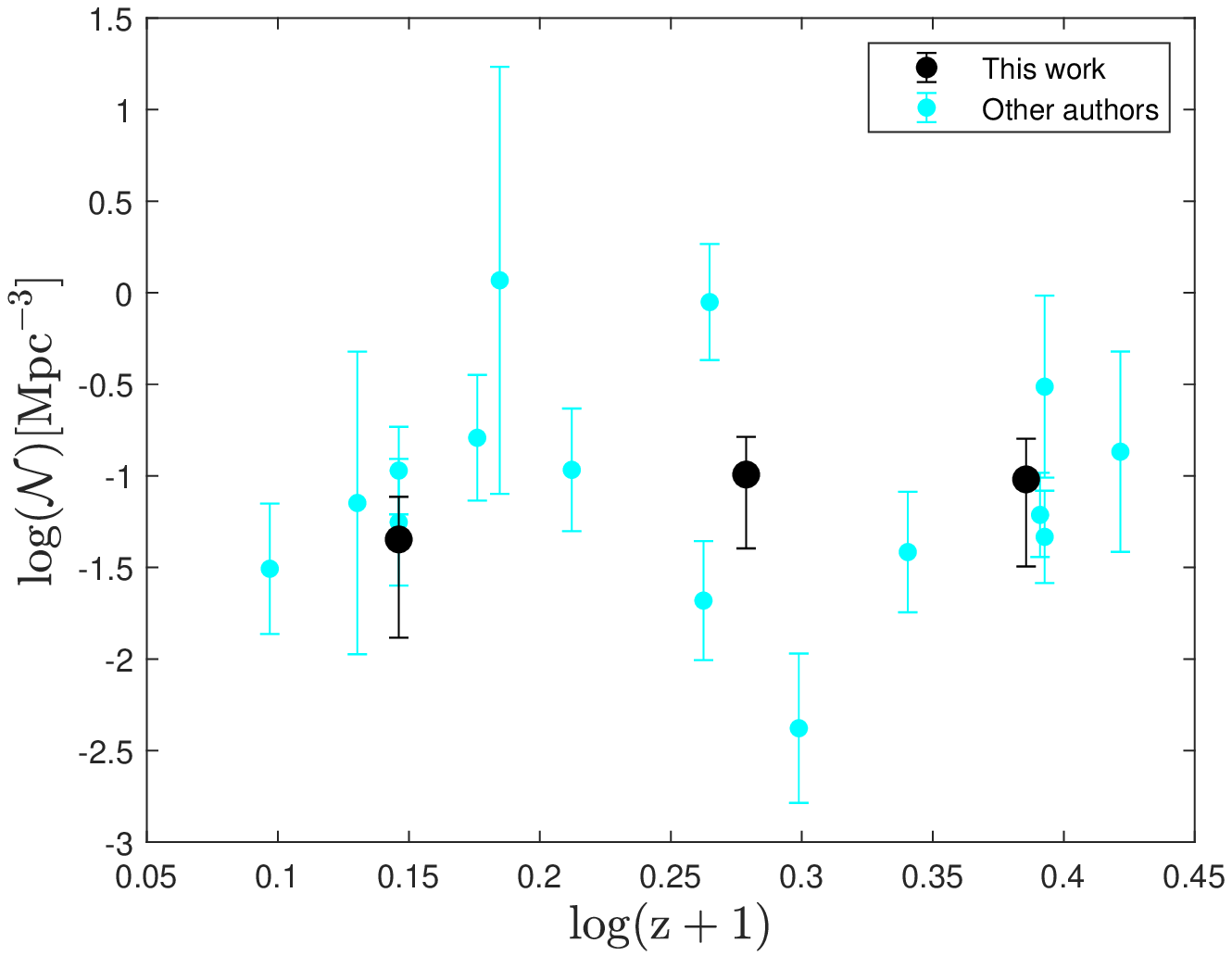}
      \caption{Number density as a function of redshift for 
      the \otelo\ emitters (black dots), and authors from 
      the literature (cyan dots). The figure includes data 
      from \cite{ly2007} (\ha  , \oiiis\ and \oii ),
      \cite{drake2013} (\ha\ and \oii ), 
      \cite{khostovan2015} (\hb ), and 
      \cite{hayashi2013} (\oii ).
      }
      \label{den1}
\end{figure}

Each LF can also be integrated to obtain the corresponding number density ($\mathcal{N}$) for each redshift sample (\ha\,, \hb\ and \oii).
\begin{equation}
    \mathcal{N}=\int_{0}^{\infty}\phi(L)\,{\rm d}L,
\end{equation}
In this case, we employed a numerical integration technique. As integration limits, we used $38<\log(L)[\mathrm{erg/s}]<43$ to cover the full range of the possible masses of the emitters. 
We obtained $\log(\mathcal{N}_{\rm {H\alpha}}[{\rm Mpc^{-3}}])=-1.34^{+0.23}_{-0.53}$, $\log(\mathcal{N}_{\rm {H\beta}}[{\rm Mpc^{-3}}])=-0.99^{+0.21}_{-0.40}$ and\\ $\log(\mathcal{N}_{\rm {[OII]}}[{\rm Mpc^{-3}}])=-1.02^{+0.22}_{-0.48}$.
In Fig. \ref{den1} we have represented the number density as a function of the logarithm of the redshift for the \ha\ , \hb\ and \oii\ bands in the \otelo\ survey (black dots). For comparison, we also have represented data from \cite{ly2007}, \cite{drake2013}, \cite{khostovan2015} and \cite{hayashi2013} (cyan dots).

The data points from previous works include galaxies not filtered by mass that are \textit{in general} more massive than those in our sample. The scatter reflects a non-conclusive trend in the evolution with redshift, especially in the range of z$\sim0.8$. Our work, based mainly on low-mass galaxies ($M_{*}<10^{10}\, {\rm M_{\odot}}$), clearly establishes a flat slope in the evolution of the number density with redshift.

\section{Discussion and conclusions}

By estimating the SFR for low-mass galaxies in the three groups of emitters from $0.4<z<1.43$ detected in the \otelo\ survey, we find little or no evolution along with this redshift range.
The different limiting magnitude achieved at each redshift window, with the lowest mass galaxies detected at $z\sim0.4$, explains the apparent lower values for the SFR and sSFR at this redshift. At the same time, the smaller cosmic volume mapped explains the lack of high mass \ha\ emitters.

Nevertheless, the data for all the emitters follow the MS defined by \cite{RP2015} for both the SFR and sSFR.

We fitted a power law to the stacked and binned sSFR relation with the stellar mass, obtaining $\beta_{\rm H\alpha}=-0.1\pm0.2$, $\beta_{\rm H\beta}=-0.5\pm0.2$ and $\beta_{\rm [OII]}=-0.6\pm0.3$. These results expand previous works \citep{ramraj2017} to masses as low as $\log (M_*/M_{\sun})=7.12$ and to redshifts up to 1.43. 
We find shallower slopes when comparing with \cite{sobral2011}. 
This flatter slope in the relationship between sSFR vs the stellar mass indicates a more constant star formation history of low mass galaxies, as suggested by \cite{ramraj2017}. 

By integrating the LFs for each group, we were able to derive the SFRD and we represented it as a function of the redshift. We find no evolution with $z$ of the SFH for our sample, formed mainly by low-mass galaxies. Even more, when integrating only objects up to  $M_{*}=10^{9}\, {\rm M_{\odot}}$ we get the same result. The evolution found in previous works \citep{ly2007,morioka2008,sobral2013,harish2020} would come from the contribution of higher mass galaxies.
The value we obtained with \ha\ data seems to be somewhat smaller than those from the other emitters. Nevertheless, its value is within the uncertainties. This may be attributed to the cosmic variance effect due to the smaller volume sampled when compared with the \hb\ and \oii\ data. These results confirm the ones obtained by \cite{davies2009} for low-mass galaxies at $z\sim1$, and extend them to lower ($z\sim0.4$) and higher ($z\sim1.43$) redshifts.

We performed a numerical integration of the number density of the \otelo\ galaxies and data in the literature. No evolution for low-mass galaxies was found either.

Taking into account all these effects, the resulting picture is a constant number density of galaxies, regardless of their masses, up to redshift $z\sim1.43$. At the same time, there is a gradual change of the relative importance of the star forming processes, from high-mass galaxies to low-mass galaxies with decreasing redshift. The contribution of low mass galaxies is constant with redshift up to $z\sim1.43$, while the contribution of high mass galaxies increases with redshift.  This may indicate the presence of the \textit{Downsizing} effect, as described by \cite{cowie1996}. 

We can conclude that the low mass galaxies are
a valid baseline comparator of the star formation activity independently of the redshift window and the mass range considered. This allows a new approach to the study of the evolution of the star formation rate of intermediate to high mass galaxies along the cosmic times.

\acknowledgments

{
The Authors thank the anonymous referee for her/his feedback and constructive suggestions, which have contributed to significantly improve the manuscript.

This  work  was  supported  by  the  Spanish  Ministry  of  Economy  and
Competitiveness  (MINECO) under  the  grants  AYA2014 -- 58861 -- C3 -- 1 -- P,
AYA2014  -- 58861  -- C3  -- 2  -- P,  AYA2014  -- 58861  -- C3  -- 3  -- P, 
AYA2013  -- 46724  -- P, AYA2017  -- 88007  -- C3  -- 1  -- P, AYA2017  -- 88007  -- C3  -- 2, MDM  -- 2017  -- 0737 (Unidad de Excelencia Mar\'ia de Maeztu, CAB).  APG acknowledge support from ESA through the Faculty of the European
Space Astronomy Centre (ESAC) - Funding reference ESAC\_549/2019.

BC wishes to thank Carlota Leal \'Alvarez by her support during the development of this paper.
EJA acknowledges funding from the State Agency for Research of the Spanish MCIU through the ``Center of Excellence Severo Ochoa" award to the Instituto de Astrof\'isica de Andaluc\'ia (SEV-2017-0709) and from grant PGC2018-095049-B-C21.

Based on observations made with the Gran Telescopio Canarias (GTC), installed in the
Spanish Observatorio del Roque de los Muchachos of the Instituto de Astrof\'isica de
Canarias, in the island of La Palma.}

{This study makes use of data from AEGIS, a multiwavelength sky survey conducted with the
Chandra, GALEX, Hubble, Keck, CFHT, MMT, Subaru, Palomar, Spitzer, VLA, and other telescopes
and supported in part by the NSF, NASA, and the STFC.}

{Based  on  observations  obtained  with  MegaPrime/MegaCam,  a  joint  project  of  CFHT  and
CEA/IRFU, at the Canada-France-Hawaii Telescope (CFHT) which is operated by the National
Research Council (NRC) of Canada, the Institut National des Science de l'Univers of the
Centre National de la Recherche Scientifique (CNRS) of France, and the University of
Hawaii.  This work is based in part on data products produced at Terapix available at
the Canadian Astronomy Data Centre as part of the Canada-France-Hawaii Telescope Legacy
Survey, a collaborative project of NRC and CNRS.}

{Based on observations obtained with WIRCam, a joint project of CFHT, Taiwan, Korea, Canada,
France, at the Canada-France-Hawaii Telescope (CFHT) which is operated by the National
Research Council (NRC) of Canada, the Institute National des Sciences de l'Univers of the
Centre National de la Recherche Scientifique of France, and the University of Hawaii.
This work is based in part on data products produced at TERAPIX, the WIRDS (WIRcam Deep
Survey) consortium, and the Canadian Astronomy Data Centre. This research was supported by
a grant from the Agence Nationale de la Recherche ANR-07-BLAN-0228}

%



\bibliographystyle{aasjournal}
\bibliography{biblio}



\end{document}